# A NOVEL TECHNIQUE OF POWER CONTROL IN MAGNETRON TRANSMITTERS FOR INTENSE ACCELERATORS*

G. Kazakevich[#], R. Johnson, M. Neubauer, Muons, Inc, Batavia, IL 60510, USA
V. Lebedev, W. Schappert, V. Yakovlev, Fermilab, Batavia, IL 60510, USA

*Abstract*

A novel concept of a high-power magnetron transmitter allowing dynamic phase and power control at the frequency of locking signal is proposed. The transmitter compensating parasitic phase and amplitude modulations inherent in Superconducting RF (SRF) cavities within closed feedback loops is intended for powering of the intensity-frontier superconducting accelerators. The concept uses magnetrons driven by a sufficient resonant (injection-locking) signal and fed by the voltage which can be below the threshold of self-excitation. This provides an extended range of power control in a single magnetron at highest efficiency minimizing the cost of RF power unit and the operation cost. Proof-of-principle of the proposed concept demonstrated in pulsed and CW regimes with 2.45 GHz, 1kW magnetrons is discussed here. A conceptual scheme of the high-power transmitter allowing the dynamic wide-band phase and mid-frequency power controls is presented and discussed.

## INTRODUCTION

SRF cavities of the modern superconducting accelerators are typically manufactured from thin sheets of niobium to allow them to be cooled at minimized power of cryo-facilities. Fluctuations of helium pressure, acoustic noise caused by liquid He flux, etc., all cause mechanical oscillations of the cavity walls changing the cavity resonant frequency. This results in parasitic amplitude and phase modulation of the accelerating field in the cavity. The cavity detuning caused by the mechanical oscillations is typically a few tens of Hz [1, 2]. At very high external quality factor of the SRF cavities, $Q_E$, the bandwidth of the fundamental mode of the RF oscillation in the cavities has the same order as the detuning value. This causes deviations of the accelerating field phase and amplitude. The parasitic modulations are not associated with instability of the RF source; they exist even if the RF source is ideally stable. Thus, only the dynamic phase and power control of the RF sources locking phase and amplitude of the RF field in the SRF cavity [3], allows keeping stable phase and amplitude of the accelerating field in the cavity. The bandwidth of the power and the phase control of the RF source is determined by necessary suppression of the amplitude and phase deviations; e.g., for suppression of the amplitude modulation to deviations level of <1% at the deviations of ~140% one needs suppression >40 dB, at the bandwidth of the control ~10 kHz.

The traditional RF sources (klystrons, IOTs, solid-state amplifiers) can provide such control; however the capital cost of the RF system of the large-scale superconducting accelerator will be a significant part of the project cost. The magnetrons controlled by the phase-modulated resonant driving RF signal may provide dynamic phase and power control with the capital cost in a few times less than the traditional RF sources [4, 5] at a higher efficiency. This will also reduce the cost of operation of the RF sources.

Two methods allowing power control in magnetron transmitters were suggested recently: using power combining from two magnetrons with a 3-dB hybrid [ibid.]; or by an additional modulation of the depth of phase-modulated signal driving a single magnetron [6]. The last one is applicable at a very high $Q_E$ value. The average relative efficiency in the range of power control of ~10 dB for the both methods is about of 50%.

We propose a novel technique of power control which keeps a wide bandwidth for the phase control and provides the range of the power control up to 10 dB by variation of current in the extended range [7]. For such a range the minimum magnetron current has to be much less than the minimum current available in free run. This is realized in the magnetron driven by a resonant wave and fed by the voltage less than threshold of self-excitation [7]. Note, that the proposed technique is applicable for any value of $Q_E$.

The proposed technique of the magnetron power control provides highest efficiency in comparison with methods described in Refs. [4-6]. This will allow significantly decrease the capital and operating costs of the ADS class projects. The proposed method increasing the transmitter efficiency at power control in magnetrons can be used in combination with methods described in Refs. [4-6] maximizing efficiency at the wideband power control.

The concept of a controllable operation of the magnetron fed by the voltage less than the threshold of self-excitation was substantiated by a developed kinetic model [7]. Demonstration of proof-of-principle of the proposed method of the wide-range power control in magnetrons is presented and discussed here.

## A WIDE-RANGE POWER CONTROL IN PRE-EXCITED CW MAGNETRONS

Proof-of-principle of the developed technique of the power variation in pre-excited magnetrons was demonstrated in experiments with 2.45 GHz, 1 kW tubes. In pulsed regime the CW magnetron type OM75P(31) with a permanent magnet was fed by a pulsed modulator at pulse duration of 1.2 ms. Before the experiments, the magnetron was in use for about of 8 years, thus it could provide output power up to ≈500 W.

* Supported by Fermi Research Alliance, LLC under Contract No. De-AC02- 07CH11359 with the United States DOE in collaboration with Muons, Inc.
[#]e-mail: gkazakevitch@yahoo.com; grigory@muonsinc.com

The experiments with power variation by the magnetron were performed with the setup [5], shown in Fig. 1.

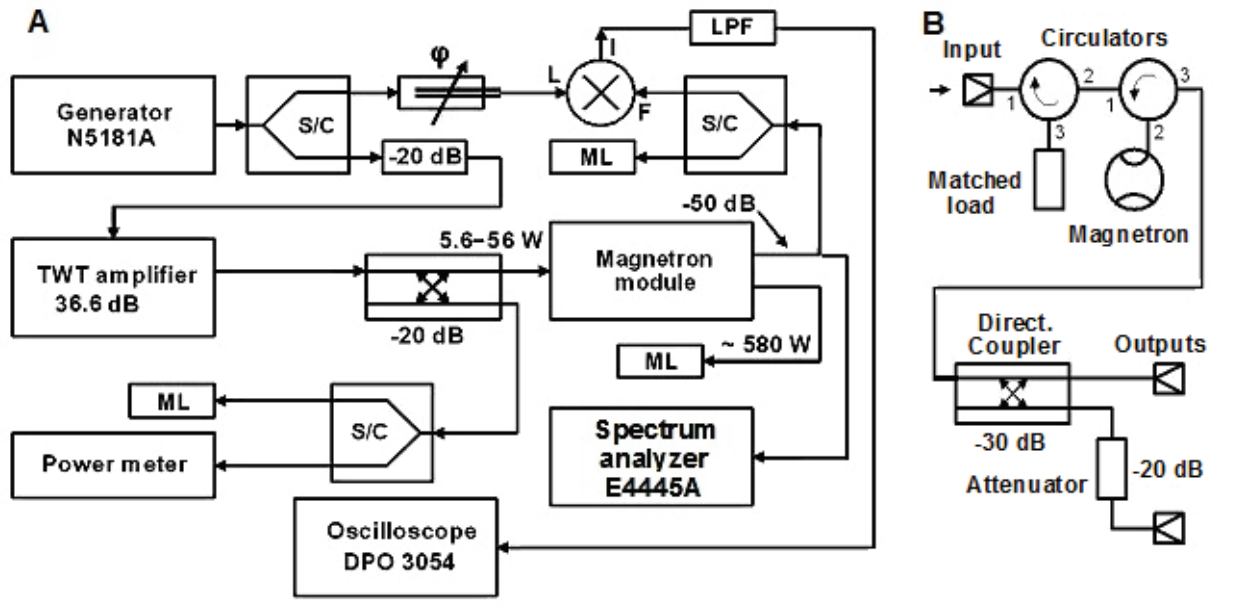

Figure 1: A- Experimental setup to measure the magnetron current, voltage, spectrum, output power, power of the pre-exciting signal and phase vs. the power variation. ML is a matched load; S/C is a 3-dB splitter/combiner. B-Scheme of the magnetron module.

The pulsed magnetron modulator was fed by a switching High Voltage (HV) stabilized power supply. When the magnetron was driven by a resonant (injection-locking) wave the measurements were performed at the frequency with -2.8 MHz offset relative to the magnetron average free run frequency at an output power of ≈450 W. The magnetron pulsed voltage and current were measured by a compensated divider and a transducer, respectively. The measurement errors do not exceed ±1%.

Measured dependence of the magnetron power on the magnetron current and Volt-Amp. (V-I) characteristics of the free running or pre-excited magnetron at various reso-nant driving signal are shown in Figs. 2, 3 [7].

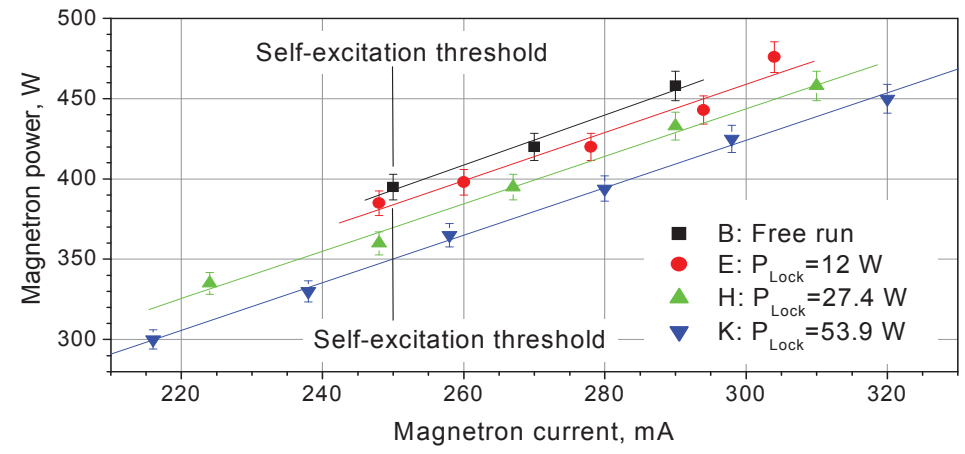

Figure 2: Dots with error bars: dependence of the magnetron pulsed power on the magnetron current for free running and pre-excited tube at various power of the pre-exciting (locking) signal. Solid lines: linear fits of the measured data.

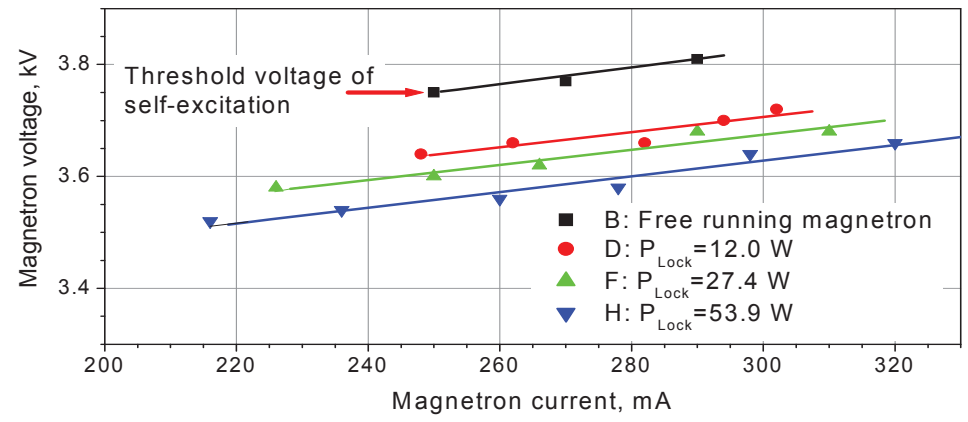

Figure 3: Measured at a constant magnetic field the V-I characteristics of the free run-ning, dots B, or pre-excited injection-locked magnetron, dots D, F, H at various $P_{Lock}$ [7]. Solid lines are linear fits.

The measured magnetron V-I characteristics show that the ratio of the magnetron dynamic impedance to the magnetron static impedance, $Z_D/Z_S$, is ~0.1. Thus, significant variations of the magnetron current (power) require insignificant variations of the magnetron feeding voltage: $\Delta P/P \approx (Z_S/Z_D) \cdot \Delta U/U$.

Measurements shown in Figs. 2, 3 demonstrate operation of the magnetron driven by a sufficient locking signal below the threshold of self-excitation in free run (Hartree voltage) and an extended range of power control.

More detailed measurements of power variation at various values of the locking power were performed in CW mode with 2.5 GHz, 1.2 kW magnetron type YJ1540 with a permanent magnet [7]. The magnetron was driven (frequency-locked) by the HP 8341A generator via a solid-state amplifier and the 36.6 dB TWT amplifier providing the CW locking power up to 100 W. The magnetron was fed by Alter switching HV power supply type SM445G, operating as a current source allowing the current control. Operation at the current less than minimum in free run (≈140 mA at the minimum magnetron power of ≈350 W) was possible at the voltage below the critical voltage in free run, Fig. 4 [7].

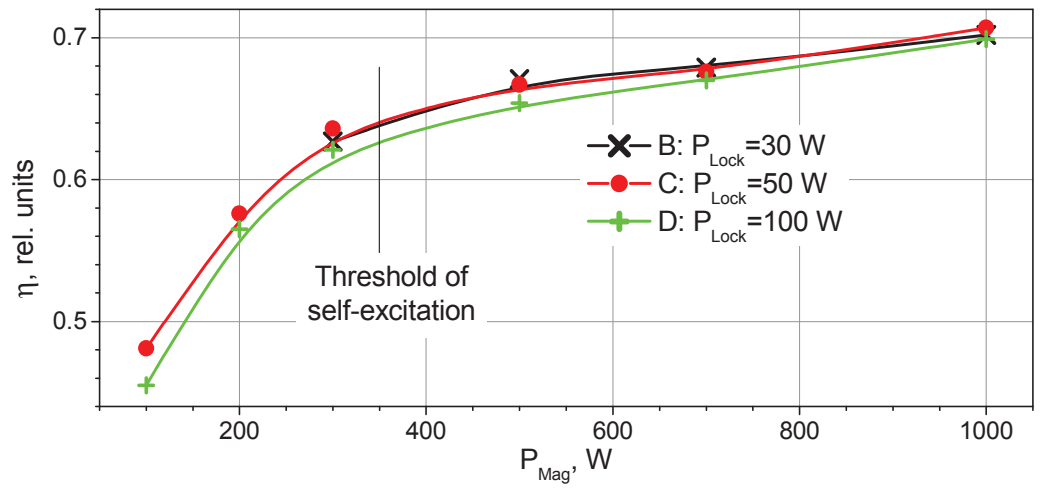

Figure 4: Measured absolute efficiency of the pre-excited injection-locked 1.2 kW magnetron with power control by management of the magnetron current in a wide range.

Magnetron efficiency vs. the range of power regulations at various methods of power control is plotted in Fig. 5.

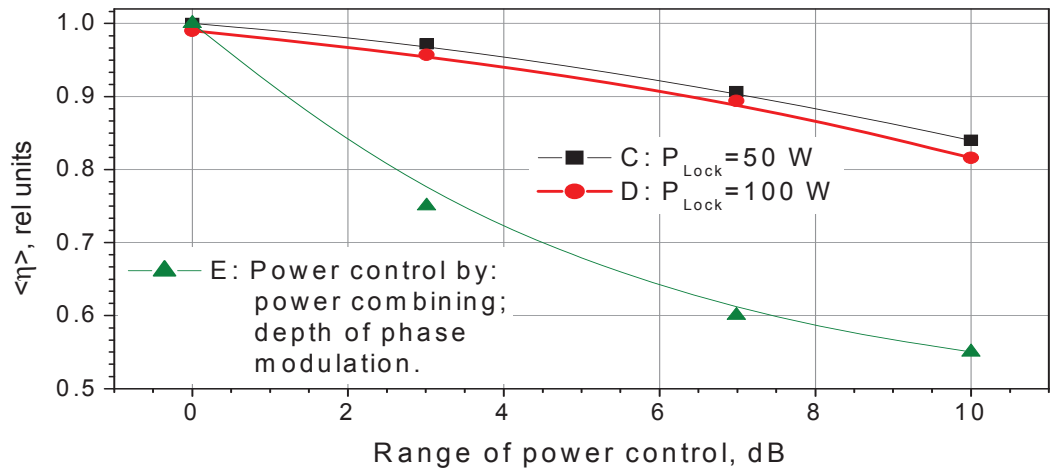

Figure 5: Relative magnetron efficiency vs. range of power variations at various methods of control. Traces C and D are average efficiency of the pre-excited injection-locked 1.2 kW magnetron measured at the extended current control. Trace E represents calculated average efficiency of 1 kW magnetrons at power combining [5], or at management of the depth of phase modulation [6].

Spectra of the carrier frequency of the injection-locked magnetron at various power levels in the range of 10 dB at the proposed power control are stable and do not demonstrate any broadening or shifts, Fig. 6 [7]. This verifies the adequacy of the proposed power control in the injection-locked magnetrons to requirements of SRF cavities.

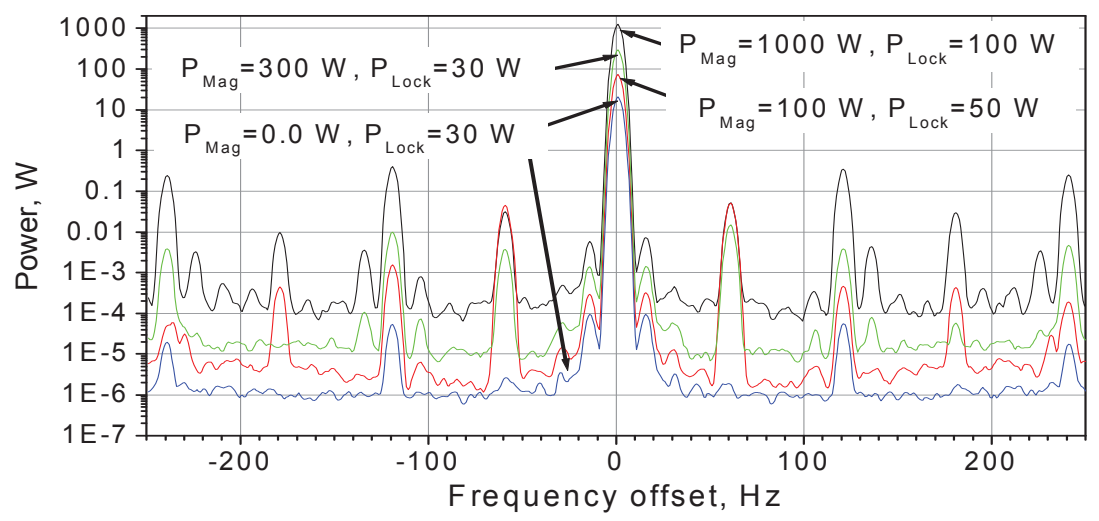

Figure 6: Offset of the carrier frequency at various powers of the magnetron, $P_{Mag}$, and the locking signal, $P_{Lock}$. Trace $P_{Mag}$=0.0 W, $P_{Lock}$=30 W shows the injection-locking signal when magnetron anode voltage was OFF.

Measured spectral density of noise of the injection-locked magnetrons in the range of output power of 10 dB did not demonstrate a notable increase at low power of the magnetron at $P_{Lock}$=100 W (-10.8 dB), Fig. 7 [7].

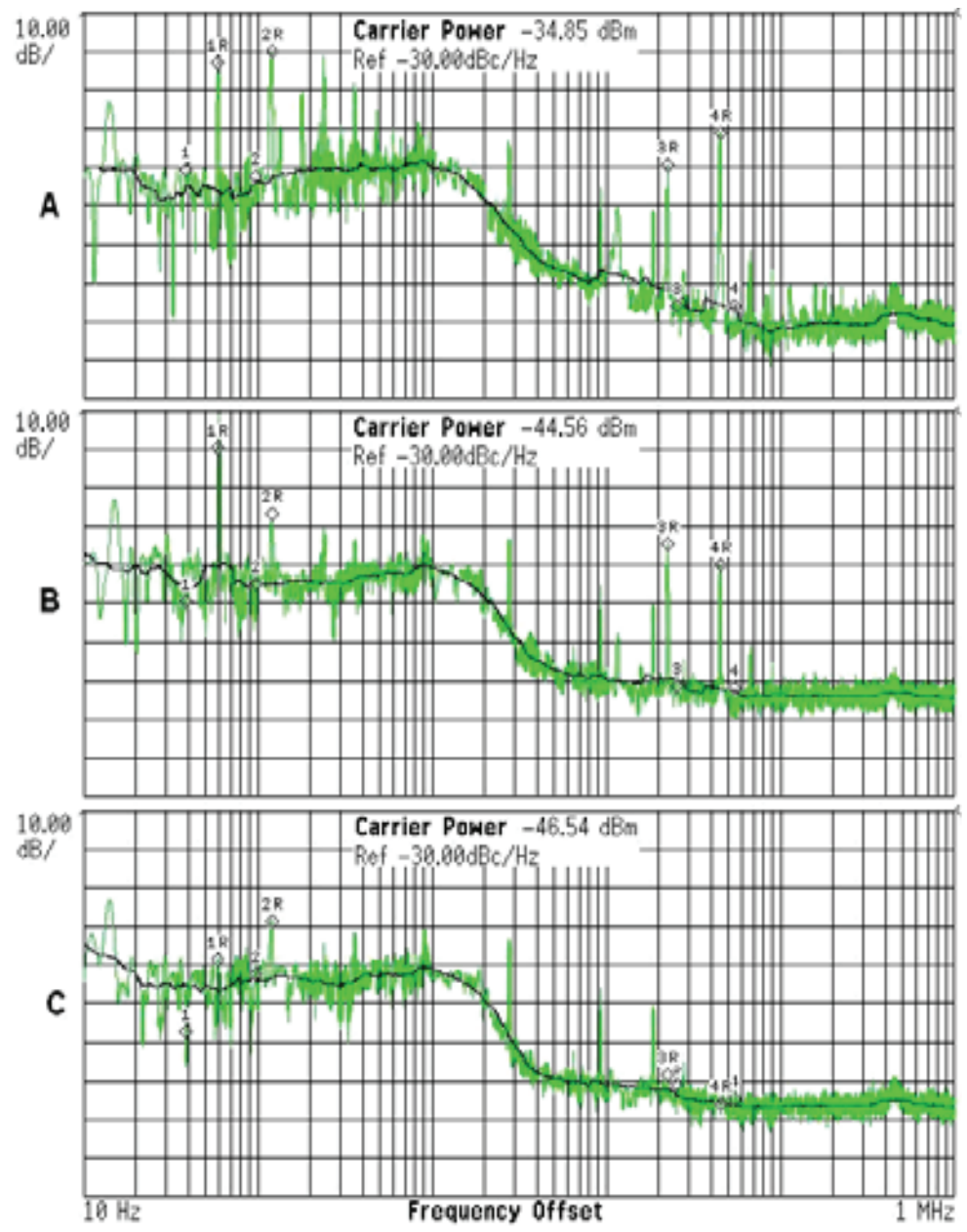

Figure 7: Spectral density of noise of the frequency-locked magnetron at output power of 1000 W and 100 W, traces A and B, respectively. Traces C are the spectral density of the injection-locking signal. The black traces show the averaged noise spectra.

Capability of the proposed method for a deep dynamic power control was verified using a modulation of the magnetron current by the switching power supply, Fig. 8.

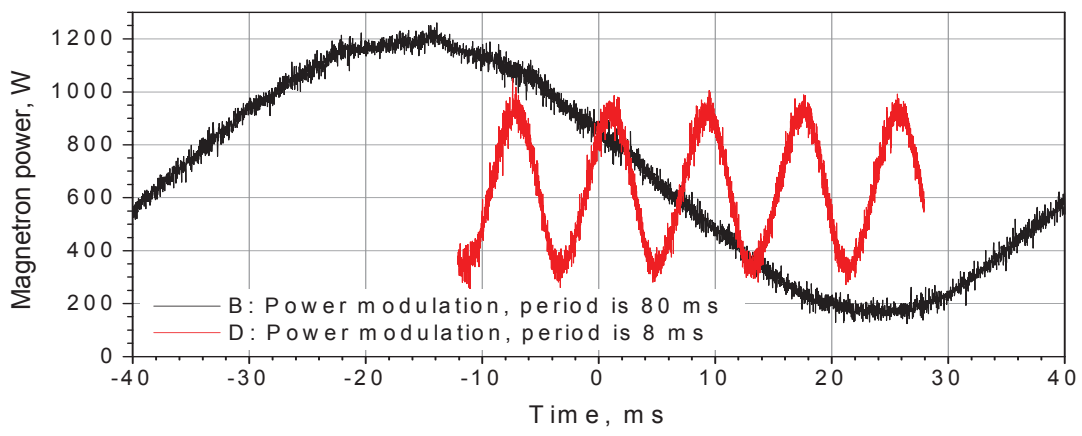

Figure 8: Modulation of the magnetron power managing the magnetron current by a harmonic signal controlling the SM445G HV switching power supply [7].

The modulation was realized by a harmonic modulation of the voltage controlling the power supply current. The traces of the magnetron current measured by a transducer, Fig. 9, represent calibrated harmonic RF power modulation. Error in the RF power calibration does not exceed 1%. The traces were averaged over 16 runs reducing the noise caused by operation of the switching power supply.

## THE HIGHLY-EFFICIENT MAGNETRON TRANSMITTER CONCEPT

The proposed method of the power control in frequency-locked magnetrons allows to simplify the conceptual scheme of the vector control in two-channel magnetron transmitter [5], utilizing an extended (up to 10 db) power control in a single 2-cascade magnetron, Fig. 9.

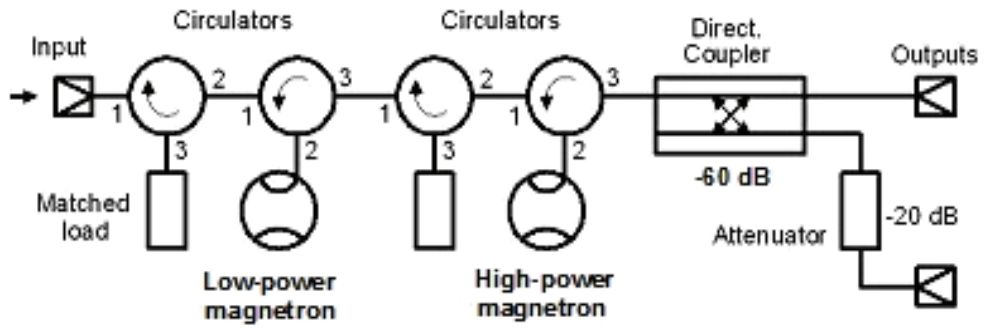

Figure 9: Conceptual scheme of a single 2-cascade magnetron allowing phase and power control.

In this scheme the first, low-power magnetron provides the phase modulation (control) of the signal frequency-locking the second, high-power magnetron. The power control in the required range (up to 10 dB) is realized by modulation (control) of current in the high-power magnetron which can operate at the voltage less than the critical voltage in free run mode. Stable and low noise operation of the high-power injection-locked magnetron at voltage less than the critical in free run is provided by pre-excitation of the tube by a sufficient magnitude of the injection-locking signal.

The control (modulation) of the magnetron current causes phase pushing in the frequency-locked magnetron. At a bandwidth of the phase control in MHz range, one expects the phase pushing elimination to a level less than -50 dB suitable for various superconducting accelerators.

The bandwidth of the proposed power control is determined by capability of control of the magnetron current in the HV power supply. Presently the bandwidth of the current control up to 10 kHz is available without efficiency compromising.

## SUMMARY

A novel method of mid-frequency dynamic power control in frequency-locked magnetrons has been proposed and verified in experiments with 2.45 GHz, 1 kW, CW tubes. The method utilizes a pre-excitation of the magnetrons with a sufficient injection-locking signal. It allows operation of the magnetron at the voltage less than the critical voltage in free run at a wideband phase control. This provides variation of the magnetron power in wide range at precisely-stable carrier frequency, low noise and highest efficiency. Thus combined with methods allowing wideband magnetron control the proposed method will provide highest efficiency for modern accelerators.